\documentclass[traditabstract]{aa}
\usepackage{txfonts}
\usepackage{natbib}
\bibpunct{(}{)}{;}{a}{}{,}
\usepackage{graphicx}
\begin{document}
\title{Light-element Abundance Variations in the Milky Way Halo}

\bigskip
\author{Sarah L. Martell \and Eva K. Grebel} 
\institute{Astronomisches Rechen-Institut\\
Zentrum f\"{u}r Astronomie der Universit\"{a}t Heidelberg\\
69120 Heidelberg, Germany\\
\email{martell@ari.uni-heidelberg.de}}
\date{Received 26 January 2010 / Accepted 18 May 2010}
\bigskip
\abstract{ 
We present evidence for the contribution of high-mass 
globular clusters to the stellar halo of the Galaxy. Using SDSS-II/SEGUE 
spectra of over 1900 G- and K-type halo giants, we identify for the first time a subset of 
stars with CN bandstrengths significantly larger, and CH bandstrengths 
lower, than the majority of halo field stars, at fixed temperature and 
metallicity. Since CN bandstrength inhomogeneity and the usual
attendant abundance variations are presently understood as a result of 
star formation in globular clusters, we interpret this subset of halo 
giants as a result of globular cluster dissolution into the Galactic
halo. We find that $2.5\%$ of our sample is CN-strong, and can infer based on recent models of globular cluster evolution that the fraction of halo field stars initially formed within globular clusters may be as large as $50\%$.}

\titlerunning{CN-Strong Halo Stars}
\authorrunning{Martell \& Grebel}
\maketitle

\section{Introduction}
Hierarchical structure formation is presently the dominant explanation
for galaxy formation, based on observed fluctuations in the cosmic
microwave background \citep{WMAP} and sophisticated numerical
simulations of their evolution to the present day (e.g., Diemand
et al. 2007\nocite{DKM07}; Springel et al. 2008\nocite{SWV08}). In
this picture, galaxies like the Milky Way are formed 
through the coalescence of multiple low-mass galaxies which develop
within a much larger dark matter halo. The initial disagreement
between the calculated mass function of dark matter subhaloes in the
simulations and the observed mass function of nearby dwarf galaxies
(e.g., Klypin et al. 1999\nocite{KKV99}; Moore
et al. 1999\nocite{MGG99}) is being addressed 
from several directions at the same time, from complex semianalytic simulations
that include star formation and feedback processes, and calculate the
chemodynamical evolution of Milky Way-like galaxies (e.g., Johnston
et al. 2008\nocite{JB08}; Tumlinson 2010\nocite{T10} and references therein) to
searches for extremely low-mass galaxies in the Local Group (e.g.,
Zucker et al. 2006a, 2006b\nocite{ZB06a}\nocite{ZB06b}; Belokurov et al. 2007\nocite{BZE07}).

The stellar halo of the Milky Way is thought to have been constructed
mostly through the early ($\simeq 10$ Gyr ago) accretion of low-mass
protogalaxies. The halo exhibits considerable substructure in density
and in velocity (e.g. Bell et al. 2008\nocite{BZB08}), and
kinematically distinct streams presently observed in the halo (e.g.,
Majewski et al. 2003\nocite{MSW03}; Duffau
et al. 2006\nocite{DZV06}; Mart{\'{\i}}nez-Delgado et al. 2007\nocite{MD07}) are interpreted as remnants of more recent
or ongoing merger activity. The ``ECHOS'' identified in \citet{SRB09} are interpreted as older substructure that has lost some spatial coherence with time. Studies of the abundance distributions and
star formation histories in nearby dwarf galaxies (e.g., Koch et
al. 2007a, 2007b, 2008a, 2008b\nocite{KG07}\nocite{KW07}\nocite{KG08}\nocite{KM08}; Kirby
et al. 2008\nocite{KSG08}; Aoki et al. 2009\nocite{AAS09}; Frebel et al. 2009\nocite{FKS09}, 2010\nocite{FSGW10}) have
shown that there is a reasonable concordance between the properties of
the present-day Milky Way halo (as characterized by, e.g., Schoerck et al. 2009\nocite{SCC09}) and the dwarf galaxies that would have
been available as stellar contributors early in Galactic history
(e.g., Font et al. 2006\nocite{FJB06}; Carollo et al. 2007\nocite{CB07}). 

However, the ongoing dissolution of globular clusters such as Palomar
5 (e.g., Odenkirchen
et al. 2001, 2002, 2003\nocite{OGR01}\nocite{OGD02}\nocite{OGD03}; Rockosi et al. 2002\nocite{ROG02}; Grillmair \& Dionatos 2006\nocite{GD06})
and NGC 5466 \citep{BEI06} implies that some fraction of halo stars
are initially formed in globular clusters. This dissolution is driven
by internal 2-body relaxation, stellar evolution processes and tidal
interactions with the Galactic potential, and can cause significant
mass loss over the lifetime of typical halo globular clusters (e.g.,
Gnedin \& Ostriker 1997\nocite{GO97}). 

One particular model of globular cluster formation, described in 
\citet{DVD08}, posits that all old halo clusters surviving to the 
present day have lost at least 90\% of their initial mass. Early 
in the development of the cluster, winds from AGB stars 
collected in the cluster center and formed a second generation of 
low-mass stars. Type Ia supernovae then caused the cluster to expand, 
and stars at large radii, mostly members of the first generation, were 
lost. In the model, the end result, $\simeq 10$ Gyr later, is a 
$10^{5} - 10^{6} M_{\sun}$ cluster with two stellar populations that 
differ slightly in age and abundance pattern.

This two-generation model was developed specifically to explain the
anomalous light-element abundance patterns observed in globular cluster stars,
specifically the presence in every old globular cluster in the Milky
Way of a subpopulation with typical Population II abundances, and a
second subgroup with the same metallicity but enhanced N, Na, and Mg
along with depleted C, O,
and Al. This abundance
bimodality has been studied extensively in globular clusters (Langer
et al. 1992\nocite{LSK92}, Kraft 1994\nocite{K94}, and Gratton
et al. 2004\nocite{GSC04} all provide thorough reviews of the topic),
and explanations for the second abundance subgroup have varied from
pollution \citep{CC98} to internal mixing \citep{L85} to enrichment of
star-forming gas by moderate- to high-mass stars (e.g., Cottrell \&
DaCosta 1981\nocite{CD81}; Yong et al. 2008\nocite{Y08}) or high-mass
binaries \citep{DMP09}. Surface
pollution of already-formed stars would result in abundance anomalies that are erased at first
dredge-up, while current models of deep mixing (e.g., Charbonnel \&
Zahn 2007\nocite{CZ07}) indicate that it
only begins to operate after first dredge-up. Both of these processes
have been effectively ruled 
out as explanations for abundance bimodality by the presence of abundance variations at all
evolutionary phases in globular clusters (e.g., Briley
et al. 2002\nocite{BCS02}; Harbeck et al. 2003\nocite{HSG03a}). In the primordial enrichment scenario, there is
ongoing discussion over the exact source of enriching material. In
some models the source is moderately high-mass ($\sim 4-5M_{\sun}$)
AGB stars (e.g., Parmentier et al. 1999\nocite{PJM99}), while
\citet{DCM07} claim that rotating high-mass ($M \ga 10 M_{\sun}$)
stars are a better source for processed material because of their very
short lifetimes and \citet{DMP09} prefer high-mass binaries because of
their potentially strong mass loss and low wind velocities. 

Stars with these light-element abundance anomalies are readily
identified through strong UV/blue CN molecular absorption and
relatively weak absorption in the CH G band, and are
hereafter called ``CN-strong stars'', with the understanding that the
full abundance pattern from C through Al necessarily follows the CN
variation. They are not observed to exist in open clusters (e.g.,
Smith \& Norris 1984\nocite{SN84}; Jacobson
et al. 2008\nocite{JFP08}; Martell \& Smith 2009\nocite{MS09}) or the
halo field \citep{GSC04}. This feedback process apparently only occurs
in the high-density environment of globular clusters, and as a result
the characteristic sawtooth abundance pattern from carbon through aluminium can be used as a marker
of globular cluster origin.  

Given the contributions globular clusters are presently making to the
halo field, and the significant mass loss predicted theoretically over
the lifetime of the Galactic globular cluster system (e.g., Baumgardt
et al. 2008\nocite{BKP08}), it is intriguing that no CN-strong stars
have to date been observed in the halo. We interpret this as a
qualitative sign that the contributions to the halo field of globular
clusters as we know them today are relatively minor, as is also
suggested in \citet{Y08}. 

To test this interpretation, we searched for CN-strong halo giants in the 
Sloan Extension for Galactic Understanding and Exploration (SEGUE) survey 
\citep{Y09}. The SEGUE survey is a spectroscopic extension of imaging taken
during the Sloan Digital Sky Survey \citep{SDSS00}, with targets
selected to address questions of halo substructure and Galactic
formation history. Data were taken from 2005 August through 2008 July
using a 640-fiber multiobject spectrograph and the same telescope at
Apache Point Observatory that was used for SDSS imaging. The first
portion of the SEGUE data was made publically available in 2008 as
part of SDSS Data Release 7 (DR7), including roughly 240,000 spectra
in 200 ``pencil beam'' lines of sight containing stars chosen for
specific purposes (i.e., fields in the Sagittarius stream, M dwarfs to
study extremely local kinematic substructure, G and K giants to study
the distant halo). In addition to flux-calibrated spectra, DR7 also
offers the products of the ``SEGUE Stellar Parameters Pipeline''
(SSPP), which include derived stellar parameters like effective
temperature, [Fe/H] metallicity, and radial velocity, determined
automatically through template matching, $\chi^{2}$ minimization,
cross-correlation or grid-matching methods, as appropriate. Lee et
al. (2008a\nocite{L08a}; 2008b\nocite{L08b}) and \citet{AP08} give
thorough explanations of the SSPP pipeline and process.

\section{The Data Set}
Data were obtained from SDSS DR7, through the online Catalog Archive Server. We selected all
SEGUE plates, and from those all stars with [Fe/H]$\leq -1.0$, $\log(\mathrm{g})
\leq 4.0$, $(g-r)_{\mathrm{0}} \geq 0.2$, and mean signal-to-noise per pixel (SNR)
larger than 20 were chosen. Additionally, the errors on various
derived parameters were required to be small: $\sigma_{\log(\mathrm{g})} \leq
0.5$, $\sigma_{\mathrm{[Fe/H]}} \leq 0.5$ with at least three independent
[Fe/H] determinations, and reduced $\chi^{2}$ of the best-fit template
spectrum less than 2.0 both in the region of the Ca II H and K lines
and in the CH G band. 

\begin{figure} 
%\begin{center}
\resizebox{\hsize}{!}{\includegraphics{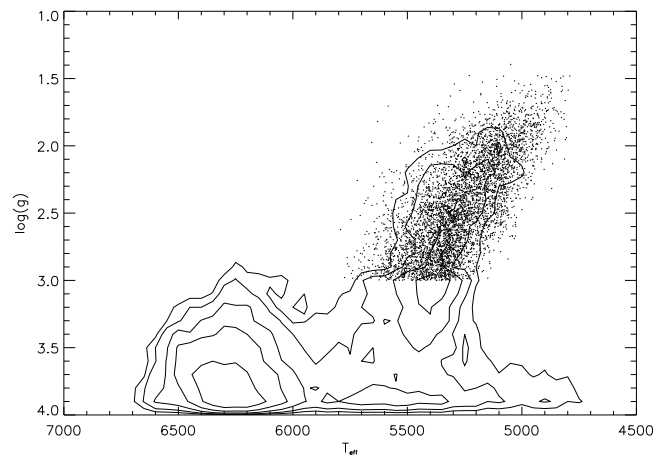}}
%\plotone{figures/f1.eps}
%\end{center} 
\caption[H-R diagram of initial and intermediate data sets]{
Hertzsprung-Russell diagram for the initial SEGUE data set, split into two subgroups. 5066 G and K giants are shown as small points, while density contours for the remaining 17718 stars rejected based on ($T_{\mathrm{eff}}$, $\log(\mathrm{g})$, [Fe/H]) position are shown as solid lines.
}
\label{ff1}
\end{figure}

This intentionally generous set of selection criteria then had to be
further sub-selected to isolate halo giants. To accomplish this
sub-selection, we lowered the limit on $\log(\mathrm{g})$ to $3.0$,
then divided the initial data set into 0.2-dex-wide bins
in [Fe/H], and removed AGB, main-sequence and turnoff stars from the
H-R diagram of each of those subsets by rejecting all points more than $3\sigma$ in $(g-r)_{\mathrm{0}}$ from the fiducial sequence. We also
made the SNR requirement more stringent, requiring that the mean SNR
per pixel in the wavelength range $4000 \leq \lambda \leq 4100$ be
larger than 15. This left 5066 halo giants, out of the original 22,784
stars. In Fig. \ref{ff1}, the small points represent those halo giants,
while density contours of stars rejected based on CMD position are
shown as solid lines. 

To facilitate later analysis, we convert the dereddened apparent
$r_{\mathrm{0}}$ magnitudes given in the SEGUE data to absolute $M_{\mathrm{r}}$
magnitudes through a simple photometric parallax calculation. We
created a grid of 10 Gyr Padova isochrones \citep{MGB08} with
metallicities ranging from [Fe/H]$=-1.0$ to [Fe/H]$=-1.8$ (the lowest
metallicity available) at a spacing
of $0.1$ dex, and interpolate between those to match the metallicity
of each individual star. The absolute $M_{\mathrm{r}}$ magnitude corresponding
to the observed $(g-r)_{\mathrm{0}}$ color of the star, on the interpolated
isochrone, is then assigned as the true absolute $M_{\mathrm{r}}$ magnitude. 

Errors in $M_{\mathrm{r}}$ were calculated by Monte Carlo 
sampling of uncorrelated errors in SSPP metallicity, $g_{\mathrm{0}}$ magnitude 
and $r_{\mathrm{0}}$ magnitude, drawn randomly from a Gaussian with a width 
equal to the reported errors on those quantities. This addition of error was done $10^{4}$ times for each star, and we take the standard deviation in those $10^{4}$ determinations of $M_{\mathrm{r}}$ as the error on $M_{\mathrm{r}}$. This error has a typical value of $0.3$ magnitudes, with the largest values ($\ga 0.5$ magnitudes) on the $3\%$ of stars with the largest errors in apparent $r_{\mathrm{0}}$ and $g_{\mathrm{0}}$. The age of the isochrones used had minimal effects on the derived $M_{\mathrm{r}}$ values, with a change of only $\pm 0.06$ magnitudes for a shift of $\mp 1$ Gyr. The limited metallicity range of the isochrones is sufficient for the purposes of the analysis in Section 3, but lower-metallicity isochrones would allow us to study the distance distribution of the roughly 1/3 of our final dataset at lower metallicity, to compare the spatial distribution of our final data set to the inner and outer halos identified in \citet{CB07}.

\begin{figure} 
%\begin{center}
\resizebox{\hsize}{!}{\includegraphics{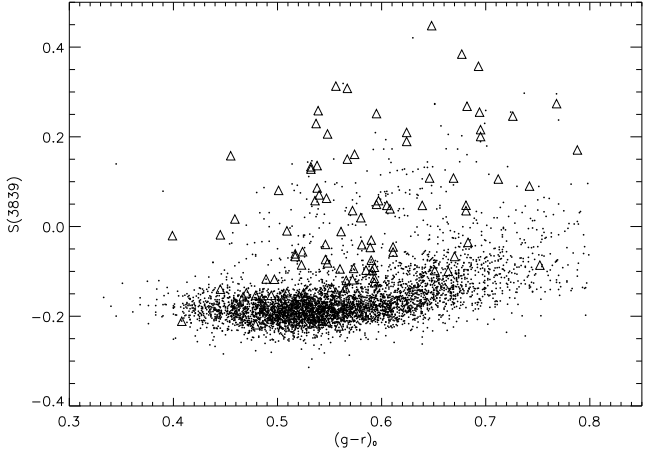}}
%\plotone{figures/f2.eps}
%\end{center} 
\caption[$S(3839)$ versus $(g-r)_{\mathrm{0}}$]{
CN bandstrength index $S(3839)$ versus dereddened $(g-r)$ color for
the G and K giants indicated in Fig. \ref{ff1}. As expected, the stars are
mainly CN-weak, and there is a slight temperature dependence. As
mentioned in the text, the stars with dramatically high $S(3839)$ are
mainly CEMP and carbon stars (shown as open triangles).
}
\label{ff2}
\end{figure}

\begin{figure} 
%\begin{center}
\resizebox{\hsize}{!}{\includegraphics{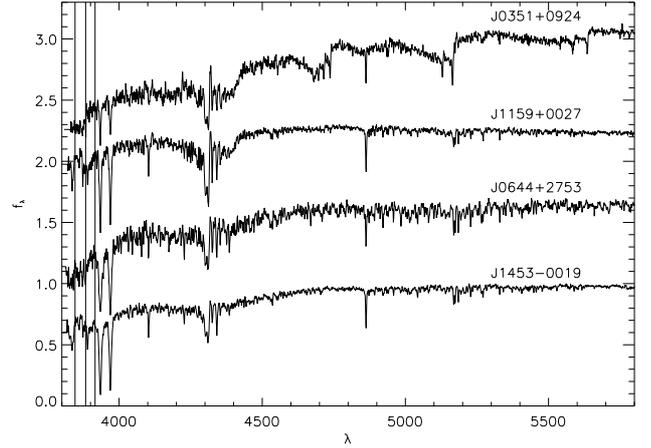}}
%\plotone{figures/f3.eps}
%\end{center} 
\caption[Sample spectra]{
Sample spectra from the four main types of star comprising our G and K giant data set: the topmost is from a carbon star, with clear CH and C$_{2}$ bands, and the next is a possible CEMP star, with a metallicity of [Fe/H]$= -2.2$ and strong CH absorption. The next spectrum down is from a CN-strong star, with stronger absorption in the $3883\hbox{\AA}$ CN band than the lowest spectrum, which shows the weak UV CN absorption seen in $98\%$ of our halo giant sample. The vertical lines mark the edges of the $S(3839)$ bands: feature ($3846\hbox{\AA}$ to $3883\hbox{\AA}$) and continuum ($3883\hbox{\AA}$ to $3916\hbox{\AA}$).
}
\label{ff3}
\end{figure}

We measured $S(3839)$ \citep{N81}, a bandstrength index for the CN
band at $3883 \hbox{\AA}$, for all halo giant spectra. $S(3839)$ measures 
the magnitude difference between the integrated flux in the CN feature and 
the integrated flux in a nearby continuum band, with more absorption in 
the feature resulting in larger bandstrength. As can be
seen in Fig. \ref{ff2}, there is a strong concentration at low
$S(3839)$ in our data set. This is to be expected, since the halo is 
primarily composed of CN-weak stars with typical Pop. II abundances. There 
is also a clear trend with temperature, in the sense that cooler stars
have larger CN bandstrengths. The cooling of stars as they ascend the 
giant branch reddens the spectra and permits more CN molecule formation, 
both of which increase the flux difference between the feature and 
continuum bands of $S(3839)$. There are, however, interesting outliers in 
Fig. \ref{ff2}: many of the stars with dramatically large $S(3839)$ are carbon 
stars (shown as open triangles), with correspondingly large CH and C$_{2}$
bandstrengths. Figure \ref{ff3} shows four sample spectra from the halo giant 
data set: the uppermost spectrum (of SDSS
J035123.90+092451.3) is a low-metallicity carbon star, the next (of
SDSS J115934.87+002748.0) is a possible CEMP star (Carbon-Enhanced
Metal-Poor, having [Fe/H]$\la -2.0$ and [C/Fe]$\ga +1.0$, and
described in Lucatello et al. 2006\nocite{LBC06} and references therein), with strong CH
absorption redward of the G band and a low metallicity, the next
spectrum (of SDSS J064411.96+275351.6) is not a carbon star,
  but is CN-strong,
and the lowest spectrum (of SDSS J145301.24-001954.1) is a typical
CN-weak halo star. 

The broad molecular features in the example
carbon-star spectrum are quite clear. We use the strength of the CH
feature around $4350\hbox{\AA}$ and the Swan (1,0) C$_{2}$ band at $4737\hbox{\AA}$ to identify carbon and CEMP stars and remove them
from the final data set. Specifically, we measure these indices:

\begin{eqnarray*}
s(c0)&=&\frac{\int_{4370}^{4400}f_{\lambda}~\mathrm{d}\lambda}{\int_{4330}^{4335}f_{\lambda}~\mathrm{d}\lambda
  +\int_{4440}^{4460}f_{\lambda}~\mathrm{d}\lambda}\\
s(c1)&=&\frac{\int_{4660}^{4742}f_{\lambda}~\mathrm{d}\lambda}{\int_{4585}^{4620}f_{\lambda}~\mathrm{d}\lambda
  +\int_{4742}^{4800}f_{\lambda}~\mathrm{d}\lambda}
\end{eqnarray*}
and consider all stars with [Fe/H]$\le -1.8$ and $s(c0)\ge -0.093$;
$-1.8 \le $[Fe/H]$\le -1.4$, $s(c0) \ge -0.05$, and $s(c1)\ge 0.15$; and [Fe/H]$\ge -1.4$, $s(c0)\ge -0.02$, and $s(c1)\ge 0.158$ to have ``strong carbon features''.
Although these carbon and CEMP stars tend to have strong UV CN bands,
as is demonstrated in, e.g., \citet{ANR02}, that is a result of
the unusually large ratio of carbon to oxygen in
their atmospheres, and is not a result of the anticorrelated C-N
abundance pattern observed in CN-strong globular cluster stars. There
are 109 stars removed from the data set because of strong carbon
features, but CN-strong stars are not entirely eliminated: carbon and
CEMP stars are shown as open triangles in Fig. \ref{ff2},
and there are clearly other stars with stronger $S(3839)$ than the
main group, at fixed $(g-r)_{\mathrm{0}}$ color. Of the 109 stars we identify as 
carbon or CEMP stars, 89 have
metallicities below [Fe/H]$=-1.8$, 11 have metallicities in the range
$-1.8\le $[Fe/H]$\le -1.4$, and 9 have a metallicity above
[Fe/H]$=-1.4$. This is not a surprising metallicity distribution:
many studies, including, e.g., \citet{G99}, \citet{LBC06} and \citet{ABS08}, have found that the
proportion of carbon stars rises at low metallicity.

\section{Distribution of CN and CH bandstrengths}
\begin{figure} 
%\begin{center}
\resizebox{\hsize}{!}{\includegraphics{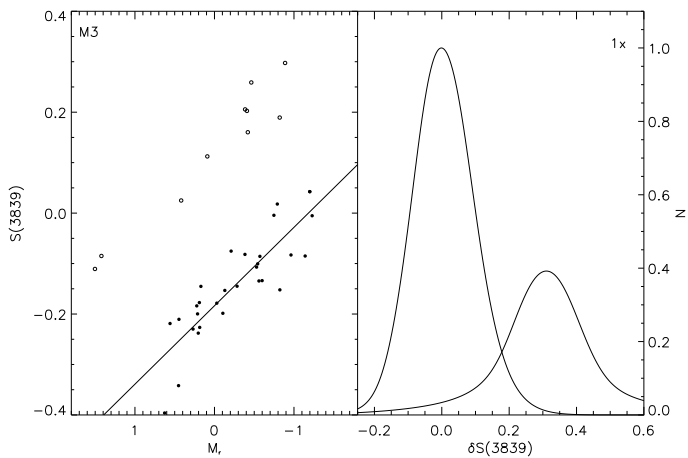}}
%\plotone{figures/f4.eps}
%\end{center} 
\caption[S(3839) distribution in M3]{
$S(3839)$ versus $M_{\mathrm{r}}$ and a generalized histogram of $\delta S(3839)$ for M3 red giants observed by SEGUE. The classic bimodal CN distribution is apparent.
}
\label{ff4}
\end{figure}

Since the purpose of this study is identifying stars in the halo with
the CN-strong chemical signature associated with globular clusters,
and our data source is the moderate-resolution spectroscopy provided
by SEGUE, our methods follow closely the techniques used in
low-resolution spectroscopic studies of CN bandstrength behavior in
globular clusters. For example, the spectral feature used is the same
$3883\hbox{\AA}$ CN band, and the index used to measure bandstrength is the
well-known $S(3839)$. One first check of this approach is
comparing CN behavior in well-studied globular clusters to halo stars
with similar metallicities observed as part of SEGUE. Figure \ref{ff4} shows
two views of the CN bandstrength data for M3, measured from SEGUE spectra of M3 giants (membership information from J. Smolinski, private communication). The left panel shows
$S(3839)$ versus absolute $M_{\mathrm{r}}$ magnitude, and the typical globular
cluster pattern is readily visible: the $S(3839)$ distribution
separates into two parallel loci, rising with increasing
luminosity. CN-strong stars (shown as open circles) are clearly distinct from CN-weak stars (filled circles) at fixed luminosity.

The right panel of Fig. \ref{ff4} converts the raw $S(3839)$ data into a
generalized histogram. In order to remove the temperature-related trend visible in
the left panel, we fit a line to the CN-weak locus and measure the
quantity $\delta S(3839)$, the vertical difference in $S(3839)$
between every point and the baseline. We then draw a generalized
histogram by representing each point as a Gaussian centered at $\delta
S(3839)$, with a FWHM equal to the error on $\delta S(3839)$, and
summing the individual Gaussians together. In Fig. \ref{ff4}, we base the Gaussian widths in the generalized histogram on the actual measurement errors on $S(3839)$, as determined by Monte Carlo sampling of the spectral error vectors reported by SEGUE. Typical values of $\sigma S(3839)$ are $0.014$ magnitudes, and we amplify that error by a factor of four in the construction of the generalized histogram. This allows for errors not accounted for in our Monte Carlo noise sampling, particularly errors in flux calibration and the SSPP-derived parameters. We calculate generalized histograms for the CN-strong and CN-weak stars independently, and
normalize both curves to the peak of the CN-weak curve, so that the
relative heights represent the relative numbers of stars in the two
groups. 

\begin{figure} 
%\begin{center}
\resizebox{\hsize}{!}{\includegraphics{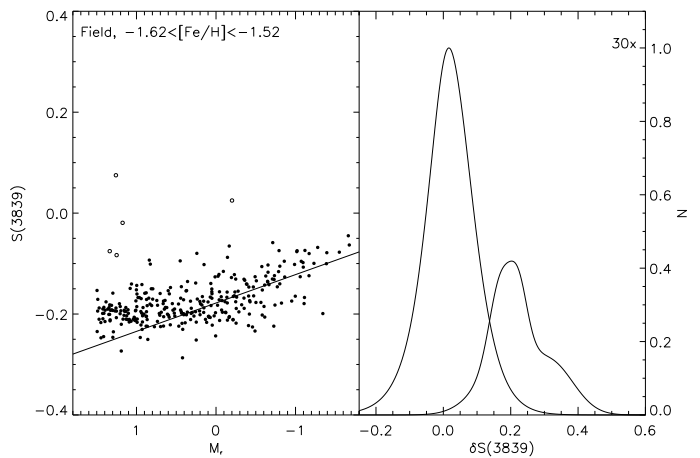}}
%\plotone{figures/f5.eps}
%\end{center} 
\caption[fmatch S(3839)]{
$S(3839)$ versus $M_{\mathrm{r}}$ and a generalized histogram of $\delta
S(3839)$ for field red giants with metallicities similar to M3 ([Fe/H] between $-1.62$ and $-1.52$). Although CN-weak stars comprise a much larger fraction of the overall data set, there are still relatively CN-strong stars seen. The CN-strong curve has been magnified by a factor of 30 for clarity.
}
\label{ff5}
\end{figure}

Figure \ref{ff5} shows analogous data for field stars in the final data set
with [Fe/H] within $\pm 0.05$ dex of M3, and similar structure can be
seen to Fig. \ref{ff4}, with interesting differences in scale. As in Fig.
\ref{ff4}, we fit a baseline to the CN-weak stars in the $M_{\mathrm{r}} - S(3839)$
plane (left panel), measure the quantity $\delta S(3839)$, and convert
that into a generalized histogram in the right panel. Errors in $\delta S(3839)$ were calculated as in Fig. \ref{ff4}. In this instance, we divide the CN-weak from CN-strong stars by shifting the baseline vertically until it encounters a significant gap. As in Fig. \ref{ff4}, CN-weak stars are shown as filled circles, and CN-strong stars as open circles. Although the two
peaks in the right panel have roughly the same separation as in Fig.
\ref{ff4}, the CN-weak group is a much more dominant component of the overall
population. We have magnified the CN-strong curve in the right panel
by a factor of 30 (noted in the upper right corner of the panel) so
that it is more easily comparable to the CN-weak curve.

It must be noted that the slope of the baseline in Fig. \ref{ff5} is
different from that in Fig. \ref{ff4}. This is unexpected, since (at
fixed metallicity) the same processes ought to be moderating the
progressive increase in CN bandstrength with increasing giant-branch
luminosity: a declining temperature that both shifts flux from the
science band of $S(3839)$ into the continuum band and permits more CN
molecule formation, along with the ``canonical extra mixing'' decribed
in \citet{DV03} that progressively depletes carbon and enhances
nitrogen abundance in the photosphere. We attribute the difference in
baseline slopes to the different mass distributions of the two
samples. Specifically, the M3 giants are an old, single-age
population, with masses all around $0.8 M_{\odot}$, while the field
giants have the possibility of being considerably younger, and
therefore more massive. A younger and more massive red giant, at fixed
luminosity, will undergo more rapid evolution, leaving less time for
deep mixing and surface abundance changes.   Indeed, as is discussed
in \citet{G89}, stars with masses greater than $2.2 M_{\odot}$ evolve
along the giant branch too quickly to even begin deep mixing. As a
result, younger giants in the field will have weaker CN bands at a
fixed $M_{\mathrm{r}}$, with the effect of pulling down the overall
baseline slope. 

\begin{figure} 
%\begin{center}
\resizebox{\hsize}{!}{\includegraphics{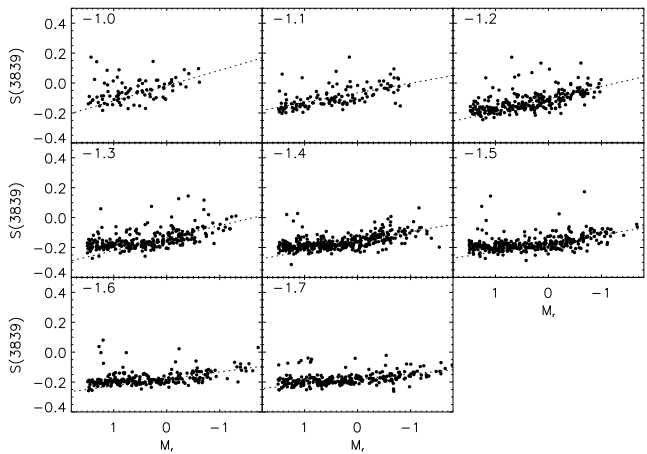}}
%\plotone{figures/f6.eps}
%\end{center} 
\caption[$S(3839)$ vs $M_{\mathrm{r}}$]{
Raw $S(3839)$ vs $M_{\mathrm{r}}$ data for the eight 0.1-dex-wide metallicity
bins in our sample. The baselines against which $\delta S(3839)$ is
measured are shown as dotted lines, and the maximum metallicity for each panel is given in the upper left corner.
}
\label{ff6}
\end{figure}

To extend this type of analysis to the full final data set, we
must carefully isolate [C/Fe] and [N/Fe] variations from the underlying
metallicity, which strongly affects the appearance of double-metallic
absorption features like $\lambda 3883$ CN. To that end, we divide the 1958 stars in the 
final data set into 0.1-dex-wide bins in [Fe/H] and search the bins
independently for a bimodal distribution of CN bandstrength at fixed
luminosity. Since CN bandstrength variations become very small at low 
metallicity, even for significant variations in [C/Fe] and [N/Fe] 
(e.g., Martell et al. 2008a\nocite{MSB08}; Briley et al. 
1993\nocite{BSHB93}), we limit our sample to the relatively metal-rich 
end of the halo. Figure \ref{ff6} shows the raw $S(3839)$ vs $M_{\mathrm{r}}$ 
distribution for field stars in each of the metallicity bins between [Fe/H]$=-1.0$ and [Fe/H]$=-1.8$. The maximum metallicity of the bin is given in the upper left corner of each
panel. As in the left panel of Fig. \ref{ff4}, the dashed line in 
each panel is the baseline against which $\delta S(3839)$ is measured. 

There are several features of note in this figure: the slope of
the baseline flattens monotonically with dropping metallicity, from
$-0.11$ in the $-1.1 \leq$ [Fe/H] $\leq -1.0$ bin, to $-0.04$ in the
$-1.8 \leq$ [Fe/H] $\leq -1.7$ bin. It is known that deep mixing is
more efficient at low metallicity: \citet{MSB08C} find that d[C/Fe]/dM$_{V}$
is twice as large at [Fe/H]$=-2.3$ as at $-1.0$, and \citet{SM79} predict 
less-compressed hydrogen-burning shells, allowing for more penetration 
by meridional circulation currents, at lower [Fe/H]. In addition, the 
gap between the CN-strong and CN-normal groups shrinks as metallicity 
declines, an effect that occurs even without reducing the size of 
variations in [C/Fe] and [N/Fe].

\begin{figure} 
%\begin{center}
\resizebox{\hsize}{!}{\includegraphics{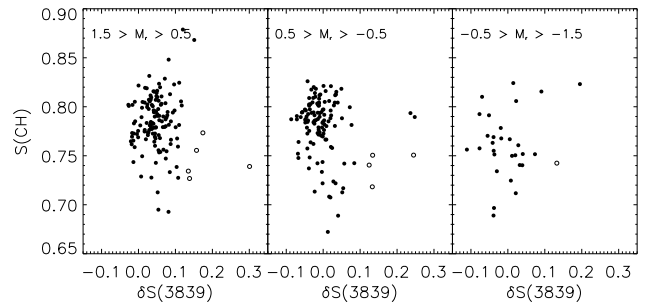}}
%\plotone{figures/f7.eps}
%\end{center} 
\caption[CN-CH, binned by luminosity]{
CN versus CH bandstrength for the $-1.4 \leq$ [Fe/H] $\leq -1.3$ metallicity bin, divided into three luminosity groups. Stars with both strong CN and relatively weak CH are shown as open circles.
}
\label{ff7}
\end{figure}

In order to identify stars with globular cluster-like carbon and
nitrogen abundances in Fig. \ref{ff6}, we look for the classic CN-CH
anticorrelation, used in globular cluster studies as a clear signal of
carbon depletion and nitrogen enrichment. Figure \ref{ff7} shows $\delta
S(3839)$ versus the CH bandstrength index $S(CH)$ \citep{MSB08b} for
the subset of the final data set with $-1.4 \leq $[Fe/H]$\leq
-1.3$. The stars are further 
subdivided by absolute $M_{\mathrm{r}}$ magnitude, since CH bandstrength is a
sensitive function of both temperature and carbon abundance, and both
decrease with rising luminosity. Stars with relatively large $\delta S(3839)$ and relatively weak $S(CH)$ in each luminosity sub-bin are shown as open circles.

\begin{figure} 
%\begin{center}
\resizebox{\hsize}{!}{\includegraphics{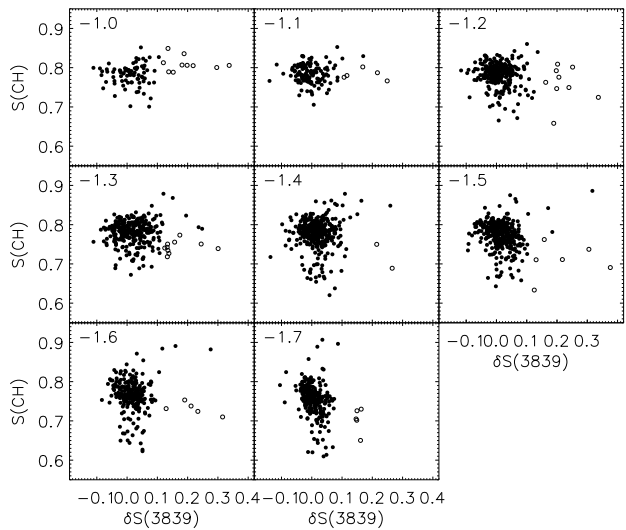}}
%\plotone{figures/f8.eps}
%\end{center} 
\caption[CN-CH, all metallicities]{
CN versus CH bandstrength for all eight metallicity bins, selected from luminosity sub-bins as in Fig. \ref{ff7}. Candidate
CN-strong stars are shown as open circles.
}
\label{ff8}
\end{figure}

We repeated this selection process, dividing the metallicity bins into luminosity sub-bins, for the other seven metallicity bins, and Fig. \ref{ff8} shows $\delta S(3839)$ versus $S(CH)$ for each of the metallicity bins, with all luminosity sub-bins collapsed together. Stars with
strong CN (and weak CH, in the lower-metallicity bins),
relative to the majority of halo field stars, are shown as open
circles in each panel. Altogether, we identify 49 stars (also shown as open
circles in Fig. \ref{ff8}) as relatively CN-strong and CH-weak. Our CH bandstrength index $S(CH)$ is calibrated based on bright red giants in the
low-metallicity globular cluster M53. It is therefore less responsive to
variations in carbon abundance in high-metallicity stars and in
fainter giants than it was designed
for. However, it was the most responsive over the full parameter range
of our data, of the nine G-band indices we considered. As a result, we
are less stringent in our CH bandstrength selection in the
higher-metallicity bins than in the lower-metallicity bins. 

\begin{figure} 
%\begin{center}
\resizebox{\hsize}{!}{\includegraphics{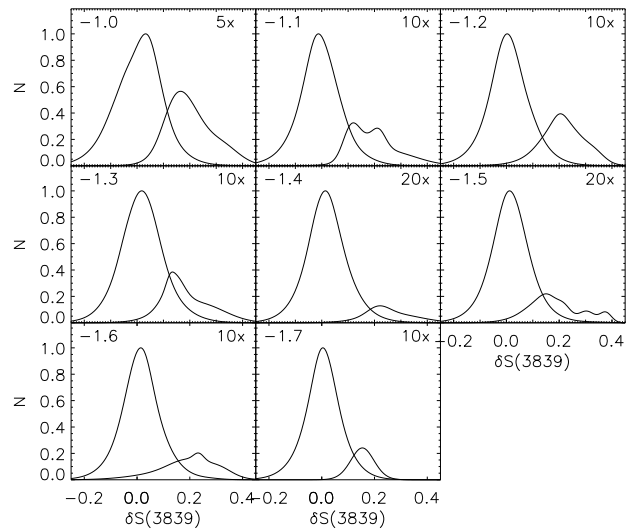}}
%\plotone{figures/f9.eps}
%\end{center} 
\caption[Generalized histograms, all metallicities]{
Generalized histograms of $\delta S(3839)$ for CN-weak and CN-strong stars in each [Fe/H] bin. As in Fig. \ref{ff6}, maximum metallicity for each panel is given in the upper left corner. As in Fig. \ref{ff5}, CN-strong histograms were amplified for clarity. The multiplicative factor for each panel is given in the upper right corner.
}
\label{ff9}
\end{figure}

The eight panels of Fig. \ref{ff9} correspond to the panels in Fig. \ref{ff6}, and
show generalized histograms of $\delta S(3839)$ for the CN-normal and CN-strong stars in
each metallicity bin, using the CN bandstrength classifications made in Fig. \ref{ff8}. As in Fig. \ref{ff5}, the CN-strong curves needed to
be amplified to be easily visible next to the CN-weak curves; the
multiplicative factor is given in the upper right corner of each
panel, and the maximum metallicity in each bin is given in the upper left corner of each panel. These generalized histograms 
are qualitatively similar to Fig. \ref{ff4}, although with a clearly different
CN-strong/CN-weak ratio: there is a separation between the two peaks
of roughly 0.2 magnitudes in $\delta S(3839)$, which shrinks as the overall
metallicity drops. 

The presence of stars in the halo field with relatively strong $\lambda 3883$ CN absorption and relatively weak absorption in the $\lambda 4320$ CH G band, at fixed metallicity and luminosity, is a strong indication that globular clusters have contributed stars to the halo field. Given current models for the origin of light-element abundance variations in globular clusters, it does not seem possible that these stars formed in the halo with these atypical [C/Fe] and [N/Fe] abundances. An investigation of the abundances of O, Na, Mg and Al, which are also known to vary in CN-strong globular cluster stars (see, e.g., Kraft 1994\nocite{K94}), would allow a more strict test, and possibly a more firm confirmation, of our claim that the 49 halo field stars we identify here as CN-strong originated in globular clusters. 

For completeness, we also mention the possibility that these stars did not originate within globular clusters, but rather that their unusual abundances are the result of mass transfer from an AGB companion in a binary system. However, as is demonstrated in \citet{LTB05}, which discusses binarity in CEMP stars, the fraction of field stars expected to be in a binary system with a companion of $3 - 8 M_{\odot}$ and orbital parameters that permit evolution of the companion up to the AGB phase, and then mass transfer but not coalescence, is quite small. There is no high-precision radial velocity monitoring program in progress for these stars; we predict that such a program would be unlikely to find binary companions.

\section{Discussion}
Although light-element abundance variations have not been observed
before in the halo, our identification of these CN-strong halo field
stars is not wholly unexpected. There are several well-understood mechanisms
for globular cluster mass loss, and theoretical studies of globular
cluster formation and evolution (e.g., D'Ercole et
al. 2008\nocite{DVD08}; Baumgardt et al. 2008\nocite{BKP08}) predict
significant mass loss in individual clusters as well as a dramatic
reshaping of the cluster mass function with time. The data set we analyzed, selected from the SEGUE survey, is
not representative of the full halo, in mass, evolutionary phase, or
metallicity. Distances to the candidate CN-strong stars range from 4
to nearly 40 kpc, with $93\%$ found within 20 kpc of the Sun. However, our result for red giants is generalizable
to all halo stars, since abundance
bimodality is observed to exist at all masses and evolutionary phases in globular
clusters. More fundamentally, all of the cuts we made in selecting the 
final data set are blind to CN and CH bandstrengths and light-element
abundances, and all nitrogen-enhanced giants with metallicities above
[Fe/H]$=-1.8$ ought to show clearly strong CN bands. Since approximately $2.5\%$ of our halo red giants exhibit
strong CN bands and weak CH bands, we expect that the same fraction of
the entire halo will contain the same abundance enhancements and
depletions. This prediction can be confirmed by observations of dwarfs
in the halo field, because main sequence stars in globular clusters
show the same abundance division as giants (e.g., Briley et
al. 2002\nocite{BCS02}). 

In order to convert $f^{\mathrm{p}}_{\mathrm{h}}$, the present-day fraction of CN-strong
halo stars, into $f^{\mathrm{gc}}_{\mathrm{h}}$, the fraction of globular
cluster-originating stars in the halo field, we must consider what
fraction CN-strong stars comprise of the stars originally formed in
globular clusters. In the two-generation scenario of \citet{DVD08},
roughly $90\%$ of stars originally formed in a globular cluster,
consisting entirely of first-generation stars (with halo-like
chemistry), are lost between the epoch of cluster formation and the
present day. This means that $f^{\mathrm{m}}_{\mathrm{gc}}$, the fraction of stars that
remain as members of the globular cluster they were formed in, is
around $0.1$. Since $f^\mathrm{p}_\mathrm{gc}$, the fraction of present-day globular
clusters stars that are CN-strong, is around $0.5$ (e.g., Kraft
1994\nocite{K94}), we can calculate that
$f^{\mathrm{gc}}_{\mathrm{h}}=\frac{f^{\mathrm{p}}_{\mathrm{h}}}{f^{\mathrm{m}}_{\mathrm{gc}}\times f^{\mathrm{p}}_{\mathrm{gc}}}$. Since
$f^{\mathrm{p}}_{\mathrm{h}}=0.025$ in the present study, this suggests that
$f^{\mathrm{gc}}_{\mathrm{h}}=\frac{0.025}{0.1\times 0.5}=0.5$, and that a remarkable
$50\%$ of the halo field originally formed in the massive star
clusters that were progenitors of the present-day globular cluster
population, with a further unknown contribution of CN-weak stars made
by globular clusters that did not survive to the present day and were
not massive enough to self-enrich.

While some numerical studies of galaxy formation (e.g., Boley et al. 2009\nocite{BLR09}) have suggested that the halo could be constructed entirely from disrupted globular clusters, there is not presently a strong consensus on the role of globular clusters in cosmological-scale galaxy formation. Precise numerical study of the dynamical evolution of globular clusters is very complicated: the number of particles is large enough, and the relevant timescales short enough, that highly accurate simulations are very time-consuming. However, the development of semianalytic prescriptions for the mass evolution of globular clusters would allow single-halo-scale simulations like those of \citet{JB08} to include them as a source of halo stars, and to predict what fraction of the halo field ought to originate in globular clusters.

\begin{acknowledgements}
SLM wishes to thank Graeme Smith and Tim Beers for helpful
conversations about this project.

Funding for the SDSS and SDSS-II has been provided by the Alfred
P. Sloan Foundation, the Participating Institutions, the National
Science Foundation, the U.S. Department of Energy, the National
Aeronautics and Space Administration, the Japanese Monbukagakusho, the
Max Planck Society, and the Higher Education Funding Council for
England. The SDSS Web Site is http://www.sdss.org/. 

The SDSS is managed by the Astrophysical Research Consortium for the
Participating Institutions. The Participating Institutions are the
American Museum of Natural History, Astrophysical Institute Potsdam,
University of Basel, University of Cambridge, Case Western Reserve
University, University of Chicago, Drexel University, Fermilab, the
Institute for Advanced Study, the Japan Participation Group, Johns
Hopkins University, the Joint Institute for Nuclear Astrophysics, the
Kavli Institute for Particle Astrophysics and Cosmology, the Korean
Scientist Group, the Chinese Academy of Sciences (LAMOST), Los Alamos
National Laboratory, the Max-Planck-Institute for Astronomy (MPIA),
the Max-Planck-Institute for Astrophysics (MPA), New Mexico State
University, Ohio State University, University of Pittsburgh,
University of Portsmouth, Princeton University, the United States
Naval Observatory, and the University of Washington.
\end{acknowledgements} 

%\bibliography{seg}

\end{document}